\documentstyle [12pt] {article}

\catcode`\@=11
\def\@citex[#1]#2{%
\if@filesw \immediate \write \@auxout {\string \citation {#2}}\fi
\@tempcntb\m@ne \let\@h@ld\relax \def\@citea{}%
\@cite{%
  \@for \@citeb:=#2\do {%
    \@ifundefined {b@\@citeb}%
      {\@h@ld\@citea\@tempcntb\m@ne{\bf ?}%
      \@warning {Citation `\@citeb ' on page \thepage \space
undefined}}%
      {\@tempcnta\@tempcntb \advance\@tempcnta\@ne%
      \@tempcntb\number\csname b@\@citeb \endcsname \relax%
      \ifnum\@tempcnta=\@tempcntb 
	\ifx\@h@ld\relax%
	  \edef \@h@ld{\@citea\csname b@\@citeb\endcsname}%
	\else%
	  \edef\@h@ld{\ifmmode{-}\else--\fi\csname
b@\@citeb\endcsname}%
	\fi%
      \else
	\@h@ld\@citea\csname b@\@citeb \endcsname%
	\let\@h@ld\relax%
      \fi}%
    \def\@citea{,\penalty\@highpenalty\,}%
  }\@h@ld
}{#1}}

\def\@citeb#1#2{{[#1]\if@tempswa , #2\fi}}
\def\@citeu#1#2{{$^{#1}$\if@tempswa , #2\fi }}
\def\@citep#1#2{{#1\if@tempswa , #2\fi}}

\def\bcites{         
	\catcode`\@=11
	\let\@cite=\@citeb
	\catcode`\@=12
}

\def\upcites{         
	\catcode`\@=11
	\let\@cite=\@citeu
	\catcode`\@=12
}

\def\plaincites{      
	\catcode`\@=11
	\let\@cite=\@citep
	\catcode`\@=12
}

\newcount\hour
\newcount\minute
\newtoks\amorpm
\hour=\time\divide\hour by 60
\minute=\time{\multiply\hour by 60 \global\advance\minute by-\hour}
\edef\standardtime{{\ifnum\hour<12 \global\amorpm={am}%
	\else\global\amorpm={pm}\advance\hour by-12 \fi
	\ifnum\hour=0 \hour=12 \fi
	\number\hour:\ifnum\minute<10
0\fi\number\minute\the\amorpm}}
\edef\militarytime{\number\hour:\ifnum\minute<10
0\fi\number\minute}

\def\draftlabel#1{{\@bsphack\if@filesw {\let\thepage\relax
   \xdef\@gtempa{\write\@auxout{\string
      \newlabel{#1}{{\@currentlabel}{\thepage}}}}}\@gtempa
   \if@nobreak \ifvmode\nobreak\fi\fi\fi\@esphack}
	\gdef\@eqnlabel{#1}}
\def\@eqnlabel{}
\def\@vacuum{}
\def\marginnote#1{}
\def\draftmarginnote#1{\marginpar{\raggedright\scriptsize\tt#1}}
\def\draft{
	\pagestyle{plain}
	\overfullrule=2pt
	\oddsidemargin -.5truein
	\def\@oddhead{\sl \phantom{\today\quad\militarytime} \hfil
	\smash{\Large\sl DRAFT} \hfil \today\quad\militarytime}
	\let\@evenhead\@oddhead
	\let\label=\draftlabel
	\let\marginnote=\draftmarginnote
	\def\ps@empty{\let\@mkboth\@gobbletwo
	\def\@oddfoot{\hfil \smash{\Large\sl DRAFT} \hfil}
	\let\@evenfoot\@oddhead}

\def\@eqnnum{(\theequation)\rlap{\kern\marginparsep\tt\@eqnlabel}%
	\global\let\@eqnlabel\@vacuum}  }

\def\blackfonts{
	\font\blackboard=msbm10 scaled\magstep1
	\font\blackboards=msbm8
	\font\blackboardss=msbm6
}

\def\nblack{            
	\def\ZZ{{Z \n{10} Z}}
	\def\NN{{N \n{14} N}}
	\def\CC{{C \n{11} C}}
	\def\RR{{R \n{11} R}}
	\def\QQ{{Q \n{12} Q}}
	\def\PP{{P \n{11} P}}
}

\def\prep{         
	\catcode`\@=11
	\input art10.sty
	\catcode`\@=12
	
	\let\small\null
	\def\blackfonts{
		\font\blackboard=msbm10
		\font\blackboards=msbm7
		\font\blackboardss=msbm5
	}
	\let\sl\it
	\twocolumn
	\sloppy
	\voffset=-2.54truecm
	\hoffset=-2.54truecm
	\flushbottom
	\parindent 1em
	\leftmargini 2em
	\leftmarginv .5em
	\leftmarginvi .5em
	\marginparwidth 48pt
	\marginparsep 10pt
	\setlength{\columnsep}{2truecm}
	\setlength{\textwidth}{25.4truecm}
	\setlength{\textheight}{17truecm}
>	\baselineskip=16pt
	\oddsidemargin .18truein
	\evensidemargin .17truein
}

\def\eqalign#1{\null\,\vcenter{\openup\jot\m@th
  \ialign{\strut\hfil$\displaystyle{##}$&$\displaystyle{{}##}$\hfil
      \crcr#1\crcr}}\,}
\def\eqalignno#1{\displ@y \tabskip\centering
  \halign
to\displaywidth{\hfil$\@lign\displaystyle{##}$\tabskip\z@skip
    &$\@lign\displaystyle{{}##}$\hfil\tabskip\centering
    &\llap{$\@lign##$}\tabskip\z@skip\crcr
    #1\crcr}}

\def\section{\@startsection {section}{1}{\z@}{3.ex plus 1ex minus
 .2ex}{2.ex plus .2ex}{\large\bf}}
\def\subsection{\@startsection{subsection}{2}{\z@}{2.75ex plus 1ex
minus
 .2ex}{1.5ex plus .2ex}{\bf}}

\def\appendix{{\newpage\section*{Appendix}}\let\appendix\section%
	{\setcounter{section}{0}
	\gdef\thesection{\Alph{section}}}\section}

\def\abstract{\if@twocolumn
\section*{Abstract}
\else 
\begin{center}
{\bf Abstract\vspace{-.5em}\vspace{0pt}}
\end{center}
\quotation
\fi}

\catcode`\@=12

\def\ibid{{\it ibid.\/}}

\newcommand{\beq}{\begin{equation}}
\newcommand{\eeq}{\end{equation}}
\newcommand{\beqa}{\begin{eqnarray}}
\newcommand{\eeqa}{\end{eqnarray}}

\newcommand{\cN}{{\cal N}}
\newcommand{\cO}{{\cal O}}
\def\noj#1,#2,{{\bf #1} (19#2)\ }
\def\jou#1,#2,#3,{{\sl #1\/ }{\bf #2} (19#3)\ }
\def\ann#1,#2,{{\sl Ann.\ Physics\/ }{\bf #1} (19#2)\ }
\def\cmp#1,#2,{{\sl Comm.\ Math.\ Phys.\/ }{\bf #1} (19#2)\ }
\def\ma#1,#2,{{\sl Math.\ Ann.\/ }{\bf #1} (19#2)\ }
\def\jd#1,#2,{{\sl J.\ Diff.\ Geom.\/ }{\bf #1} (19#2)\ }
\def\invm#1,#2,{{\sl Invent.\ Math.\/ }{\bf #1} (19#2)\ }
\def\cq#1,#2,{{\sl Class.\ Quantum Grav.\/ }{\bf #1} (19#2)\ }
\def\cqg#1,#2,{{\sl Class.\ Quantum Grav.\/ }{\bf #1} (19#2)\ }
\def\ijmp#1,#2,{{\sl Int.\ J.\ Mod.\ Phys.\/ }{\bf A#1} (19#2)\ }
\def\jmphy#1,#2,{{\sl J.\ Geom.\ Phys.\/ }{\bf #1} (19#2)\ }
\def\jams#1,#2,{{\sl J.\ Amer.\ Math.\ Soc.\/ }{\bf #1} (19#2)\ }
\def\grg#1,#2,{{\sl Gen.\ Rel.\ Grav.\/ }{\bf #1} (19#2)\ }
\def\mpl#1,#2,{{\sl Mod.\ Phys.\ Lett.\/ }{\bf A#1} (19#2)\ }
\def\nc#1,#2,{{\sl Nuovo Cim.\/ }{\bf #1} (19#2)\ }
\def\np#1,#2,{{\sl Nucl.\ Phys.\/ }{\bf B#1} (19#2)\ }
\def\pl#1,#2,{{\sl Phys.\ Lett.\/ }{\bf #1B} (19#2)\ }
\def\pla#1,#2,{{\sl Phys.\ Lett.\/ }{\bf #1A} (19#2)\ }
\def\pr#1,#2,{{\sl Phys.\ Rev.\/ }{\bf #1} (19#2)\ }
\def\prd#1,#2,{{\sl Phys.\ Rev.\/ }{\bf D#1} (19#2)\ }
\def\prl#1,#2,{{\sl Phys.\ Rev.\ Lett.\/ }{\bf #1} (19#2)\ }
\def\prp#1,#2,{{\sl Phys.\ Rept.\/ }{\bf #1C} (19#2)\ }
\def\ptp#1,#2,{{\sl Prog.\ Theor.\ Phys.\/ }{\bf #1} (19#2)\ }
\def\ptpsup#1,#2,{{\sl Prog.\ Theor.\ Phys.\/ Suppl.\/ }{\bf #1}
(19#2)\ }
\def\rmp#1,#2,{{\sl Rev.\ Mod.\ Phys.\/ }{\bf #1} (19#2)\ }
\def\yadfiz#1,#2,#3[#4,#5]{{\sl Yad.\ Fiz.\/ }{\bf #1} (19#2) #3%
\ [{\sl Sov.\ J.\ Nucl.\ Phys.\/ }{\bf #4} (19#2) #5]}
\def\zh#1,#2,#3[#4,#5]{{\sl Zh..\ Exp.\ Theor.\ Fiz.\/ }{\bf #1}
(19#2) #3%
\ [{\sl Sov.\ Phys.\ JETP\/ }{\bf #4} (19#2) #5]}

\def\beq{\begin{equation}}
\def\eeq{\end{equation}}
\def\beqar{\begin{eqnarray}}
\def\eeqar{\end{eqnarray}}
\def\non{\nonumber}

\def\nfrac#1#2{{\displaystyle{\vphantom1\smash{\lower.5ex\hbox{\sma
ll$#1$}}%
	\over\vphantom1\smash{\raise.25ex\hbox{\small$#2$}}}}}

\def\p#1{\mskip#1mu}
\def\n#1{\mskip-#1mu}
\def\stop{\p6.}
\def\comma{\p6,}

\def\lae{\mathrel{\mathop{\smash{\lower .5 ex
\hbox{$\stackrel<\sim$}}}}}
\def\lae{\mathrel{\mathop{\smash{\lower .5 ex
\hbox{$\stackrel>\sim$}}}}}

\def\l:{\mathopen{:}\,}
\def\r:{\,\mathclose{:}}

\catcode`\@=11
\def\theequation{\arabic{equation}}
\catcode`\@=12

\nblack
\bcites

\nblack

\catcode`\@=11
\def\theequation{\thesection.\arabic{equation}}
\@addtoreset{equation}{section}
\@addtoreset{footnote}{section}
\@addtoreset{footnote}{subsection}
\catcode`\@=12

\newcommand{\beqn}{\begin{equation}}
\newcommand{\eeqn}{\end{equation}}
\newcommand{\beqnarray}{\begin{eqnarray}}
\newcommand{\eeqnarray}{\end{eqnarray}}
\newcommand {\bear} [1] {\begin {array} {#1}}
\newcommand {\ear} {\end {array}}

\newcommand {\beqarn} {\begin{eqnarray*}}
\newcommand {\eeqarn} {\end{eqnarray*}}

\newcommand {\mrm} [1] {\mathrm {#1}}

\newcommand {\myref} [1]	%
	{%
	\begin{thebibliography} {99}	%
			{#1}	%
	\end {thebibliography}}

\def\emph#1{{\em #1}}
\def\mathbf#1{{\bf #1}}
\def\mathrm#1{{\rm #1}}
\def\mathit#1{{\it #1}}

\catcode`\@=11
\@addtoreset{footnote}{section}
\catcode`\@=12

\newcounter	{exinsert}	[subsection]

\renewcommand {\theexinsert}	%
{	\thesubsection.\arabic{equation}}

\renewcommand {\theequation} {\thesection.\arabic{equation}}

\catcode`\@=11
\@addtoreset{equation}{section}
\catcode`\@=12

\newenvironment {exinsert} [1]	%
{	
	\begin {quotation}	%
	\refstepcounter {equation}	%
	{\bf {#1}} \theequation. \,\,	%
}
{	
	\end {quotation}
}

\newcommand {\bprop}
{\begin {exinsert} {Proposition}
}
\newcommand {\eprop} {\end {exinsert} }

\newcommand {\bexe} {\begin {exinsert} {Exercise}}
\newcommand {\eexe} {\end {exinsert} }

\newcommand {\bexa} {\begin {exinsert} {Example}}
\newcommand {\eexa} {\end {exinsert} }

\newcommand {\literal} [1] {{ \mbox {$\mrm {#1}$}}}

\newcommand {\grad} {\nabla}

\newcommand {\ncmd} [2] {\newcommand {#1} {#2}}

\ncmd {\vgrad} {{\vec \grad}}

\ncmd {\Mv} {{\cal M}_V}
\ncmd {\Mh} {{\cal M}_H}

\ncmd {\adj} {\literal {adj}}	
\ncmd {\Mp} {M_\literal {planck}}	

\ncmd {\x} {$\times$}

\newcommand {\btab} [3]
  {\begin{table} [htb]
   \begin{center}
   \caption {#2}	\label {tab:#1}
   \begin {tabular} {#3}
   \hline}

\newcommand {\etab}
	{  \hline   \end {tabular}
   \end{center}
  \end{table}}

\ncmd {\bR} {\textbf {R}}
\ncmd {\bC} {\textbf {C}}

\ncmd {\CQ} {\cal Q}

\def\jou#1,#2,#3,{{\sl #1\/ }{\bf #2} (19#3)\ }
\def\ibid#1,#2,{{\sl ibid.\/ }{\bf #1} (19#2)\ }
\def\am#1,#2,{{\sl Acta. Math.\/ } {\bf #1} (19#2)\ }
\def\annp#1,#2,{{\sl Ann.\ Physics\/ }{\bf #1} (19#2)\ }
\def\cmp#1,#2,{{\sl Comm.\ Math.\ Phys.\/ }{\bf #1} (19#2)\ }
\def\cqg#1,#2,{{\sl Class.\ Quantum Grav.\/ }{\bf #1} (19#2)\ }
\def\dm#1,#2,{{\sl Duke\ Math.\ J.\/ }{\bf #1} (19#2)\ }
\def\ijmpa#1,#2,{{\sl Int.\ J.\ Mod.\ Phys.\/ }{\bf A#1} (19#2)\ }
\def\invm#1,#2,{{\sl Invent.\ Math.\/ }{\bf #1} (19#2)\ }
\def\jhep#1,#2,{{\sl JHEP\/ }{\bf #1} (19#2)\ }
\def\jmp#1,#2,{{\sl J.\ Geom.\ Phys.\/ }{\bf #1} (19#2)\ }
\def\jams#1,#2,{{\sl J.\ Diff.\ Geom.\/ }{\bf #1} (19#2)\ }
\def\jdg#1,#2,{{\sl J.\ Amer.\ Math.\ Soc.\/ }{\bf #1} (19#2)\ }
\def\jpa#1,#2,{{\sl J.\ Phys.\ A.\/ }{\bf #1} (19#2)\ }
\def\grg#1,#2,{{\sl Gen.\ Rel.\ Grav.\/ }{\bf #1} (19#2)\ }
\def\mpla#1,#2,{{\sl Mod.\ Phys.\ Lett.\/ }{\bf A#1} (19#2)\ }
\def\ma#1,#2,{{\sl Math.\ Ann.\/ } {\bf #1} (19#2)\ }
\def\nc#1,#2,{{\sl Nuovo\ Cim.\/ }{\bf #1} (19#2)\ }
\def\npb#1,#2,{{\sl Nucl.\ Phys.\/ }{\bf B#1} (19#2)\ }
\def\plb#1,#2,{{\sl Phys.\ Lett.\/ }{\bf #1B} (19#2)\ }
\def\pla#1,#2,{{\sl Phys.\ Lett.\/ }{\bf #1A} (19#2)\ }
\def\pnas#1,#2,{{\sl Proc..Nat.Acad.Sci.\/ }{\bf #1} (19#2)\ }
\def\prev#1,#2,{{\sl Phys.\ Rev.\/ }{\bf #1} (19#2)\ }
\def\prd#1,#2,{{\sl Phys.\ Rev.\/ }{\bf D#1} (19#2)\ }
\def\prl#1,#2,{{\sl Phys.\ Rev.\ Lett.\/ }{\bf #1} (19#2)\ }
\def\prpt#1,#2,{{\sl Phys.\ Rept.\/ }{\bf #1C} (19#2)\ }
\def\ptp#1,#2,{{\sl Prog.\ Theor.\ Phys.\/ }{\bf #1} (19#2)\ }
\def\ptpsup#1,#2,{{\sl Prog.\ Theor.\ Phys.\/ Suppl.\/ }{\bf #1} (19#2)\ }
\def\rmp#1,#2,{{\sl Rev.\ Mod.\ Phys.\/ }{\bf #1} (19#2)\ }
\def\sm#1,#2,{{\sl Selec. Math.\/ }{\bf #1} (19#2)\ }
\def\yadfiz#1,#2,#3[#4,#5]{{\sl Yad.\ Fiz.\/ }{\bf #1} (19#2) #3%
\ [{\sl Sov.\ J.\ Nucl.\ Phys.\/ }{\bf #4} (19#2) #5]}
\def\zh#1,#2,#3[#4,#5]{{\sl Zh.\ Exp.\ Theor.\ Fiz.\/ }{\bf #1} (19#2) #3%
\ [{\sl Sov.\ Phys.\ JETP\/ }{\bf #4} (19#2) #5]}

\begin {document}


\begin{titlepage}

\begin{center}
\hfill UCB-PTH-98/30,~~
LBNL-41875,~~
NSF-ITP-98-068\\
\hfill hep-th/9806021

\vskip 1.5 cm
{\Large \bf Glueball Mass Spectrum From Supergravity}
\vskip 1 cm
{\large Csaba Cs\'aki$^{1,2}$\footnote{Research Fellow, Miller Institute for
Basic Research in Science.},
Hirosi Ooguri$^{1,2,3}$,
Yaron Oz$^{1,2,3}$ and
John Terning$^{1,2}$}\\
\vskip 1cm
{$^1$ Department of Physics,
University of California at Berkeley,\\
Berkeley, CA 94720}

{$^2$ Theoretical Physics Group, Mail Stop 50A-5101,\\
 Lawrence Berkeley National Laboratory, \\
Berkeley, CA 94720}

{$^3$ Institute for Theoretical Physics,
University of California,\\
Santa Barbara, CA 93106}

{\tt csaki@thwk5.lbl.gov, ooguri@thsrv.lbl.gov, yaronoz@thsrv.lbl.gov,
terning@alvin.lbl.gov}

\end{center}

\vskip 0.5 cm
\begin{abstract}
We calculate the spectrum of glueball masses in
non-supersymmetric Yang-Mills
theory in three and four dimensions, based on
a conjectured duality between supergravity and
large $N$ gauge theories. The glueball masses
are obtained by solving supergravity
wave equations in a black hole geometry.
We find that the mass ratios are in good numerical agreement
with the available lattice data. We also compute
the leading $(g_{YM}^2 N)^{-1}$ corrections
to the glueball masses, by taking into account
stringy corrections to the supergravity action
and to the black hole metric.  We find that
the corrections to the masses are negative and
of order $(g_{YM}^2N)^{-3/2}$. Thus for a fixed
ultraviolet cutoff the masses decrease
as we decrease the 't Hooft coupling, in accordance
with our expectation about the continuum limit of
the gauge theories.

\end{abstract}
\end{titlepage}

\section{Introduction}

Recently Maldacena formulated a conjecture \cite{mal} stating
that the large $N$ limit of the maximally supersymmetric
conformal theories in $3,4$ and $6$ dimensions are
dual to superstring/$M$ theory on AdS$_4 \times {\bf S}^7$,
AdS$_5 \times {\bf S}^5$ and AdS$_7 \times {\bf S}^4$
respectively, where AdS$_d$ is a $d$-dimensional anti-de
Sitter space. More recently Witten proposed \cite{w2}
that one can extend this
duality to non-supersymmetric theories such as pure QCD.
In this case the AdS space is replaced by
the Schwarzschild geometry describing a black hole in the
AdS space.
When the curvature of the spacetime is small compared
to the string scale and the Planck scale,
superstring/$M$ theory is well-approximated by supergravity.
It was found that the supergravity description
gives results that are in qualitative agreement with
expectations for QCD at strong coupling.
This includes the area law behavior of Wilson loops,
the relation between confinement and monopole condensation,
the existence of a mass gap for glueball states,
the behavior of Wilson loops for higher representations,
and the construction of heavy quark baryonic states
\cite{w2,BISY,rty,li,w3,go}.

In this paper, we use the supergravity description
of large $N$  gauge theories to
compute the scalar glueball mass spectrum explicitly for pure QCD$_3$
and QCD$_4$.
The glueball masses in QCD can be obtained by computing
correlation functions of
gauge invariant local operators or the Wilson loops,
and looking for
particle poles. According to the refinement of
Maldacena's conjecture given in \cite{GKP,witten-one},
correlation functions of a certain class of
local operators (chiral primary operators and their superconformal
descendants) are related at large $N$ and large $g_{YM}^2N$
to tree level amplitudes of supergravity. The
correspondence between the chiral operators and the
supergravity states has been worked out in
\cite{GKP,witten-one,kleb,gk,ho,f,ozterning}.
For example, the operator ${\rm tr} F^2$ in four dimensions
corresponds to the dilaton field of supergravity
in ten dimensions.
Therefore
the scalar glueball\footnote{In the following we will
use the notation $J^{PC}$ for the glueballs, where $J$
is the glueball spin, and $P$, $C$
refer to the parity and charge conjugation quantum numbers respectively.}
$J^{PC} = 0^{++}$ in QCD which couples to ${\rm tr} F^2$
is related to the dilaton propagating in the
black hole geometry. In particular, its mass is
computable by solving the dilaton wave equation \cite{w2}.
In \cite{go},
it was shown that the correlation function of
Wilson loops is also expressed in terms of
supergraviton exchange if the distance between the
loops becomes larger than their sizes, leading again
to the supergravity wave equation.

In this paper we will solve the wave equations
numerically to obtain the glueball masses. Since
this description preserves all the symmetries of
QCD, we can identify the spin and the other quantum
numbers of the glueballs.
 The mass ratios turn out to be
in excellent agreement with the available lattice data
in the continuum limit. This is surprising since a priori
the supergravity computations are to be compared with the
strong ultraviolet coupling limit of the gauge theory
$g_{YM}^2 N \gg 1$.

As we will see,
the supergravity computation at $g_{YM}^2N \gg 1$
gives the glueball masses
in units of the fixed ultraviolet cutoff $\Lambda_{UV}$.
For finite `t Hooft coupling $\lambda = g_{YM}^2 N$,
the glueball mass $M$ would be a function of the form,
\beq
   M^2 = f(\lambda) \Lambda_{UV}^2.
\eeq
To take the continuum limit $\Lambda_{UV} \rightarrow \infty$,
we have to simultaneously take $\lambda \rightarrow 0$
so that the right-hand side of this equation becomes
of the order of the QCD mass scale $\Lambda_{QCD}$.
This in particular requires that $f(\lambda)$ decreases
as we decrease the `t Hooft coupling $\lambda$.

We compute the leading $\lambda^{-1}$ corrections to the supergravity
computation and show that this is indeed the case.
On the superstring side, the $\lambda^{-1}$
corrections are due to the
finite string tension. The leading order string
correction to the low-energy
supergravity action was computed
in \cite{grisaru,gw}. This modifies both the
background black hole metric and the supergravity
wave equation in that background.
Recently the stringy
correction to the black hole metric
was obtained in \cite{GKT} by solving the
modified supergravity equation.
We use both this metric and the string corrected wave equation
to compute the leading $\lambda^{-1}$
corrections to the $0^{++}$ glueball masses
in QCD$_3$. We find:

\noindent
(1)  The corrections to the masses are negative and
of order $\lambda^{-3/2}$:
\beq
  f(\lambda) = c_0 + c_{1} \lambda^{-3/2} + \cdots ~,~~~
c_{1} < 0,
\eeq
for the ground state and the first 5 excited levels
of the $0^{++}$ glueball. Thus, for a fixed
ultraviolet cutoff, the masses decrease
as we decrease the 't Hooft coupling, in accordance
with the expectation about the continuum limit of
QCD.

\noindent
(2) The corrections to the ratios of the glueball masses
are relatively small compared
to the correction to each glueball mass,
suggesting that the corrections are somewhat universal
for all the glueball masses.
This may indicate that the good agreement between
the supergravity computation and the lattice
gauge theory results is not a coincidence but
is due to small $\lambda^{-1}$ corrections
to the mass ratios.

This paper is organized as follows.

\noindent
In section 2, we solve the supergravity wave equations
in the AdS$_5$ black hole geometry to obtain
glueball masses in QCD$_3$
and compare the results with lattice
computations.

\noindent
In section 3, we solve the supergravity wave equations
in the AdS$_7$ black hole geometry to obtain glueball masses
in QCD$_4$ and compare the results with lattice computations.

\noindent
In section 4, we use the string theory corrections
to the low-energy supergravity action and
to the AdS$_5$ black hole geometry to estimate corrections
to the glueball masses in QCD$_3$.

\noindent
We close the paper with a summary and discussions.

\section{Glueballs in Three Dimensions}

The ${\cal N} = 4$ superconformal $SU(N)$ gauge theory in
four dimensions is realized as a low energy effective theory
of $N$ coinciding parallel D$3$ branes.
One can construct a three-dimensional
non-supersymmetric theory \cite{w2} by compactifying this theory
on ${\bf R}^3 \times {\bf S}^1$ with
anti-periodic boundary conditions
on the fermions around the compactifying circle ${\bf S}^1$.
Supersymmetry is broken explicitly by the boundary conditions.
As the radius $R$ of the circle becomes small, the fermions
decouple from the system since there are no zero frequency
Matsubara modes.
The scalar fields in the 4D theory will acquire
masses at one-loop, since  supersymmetry
is broken, and these masses become infinite as $R \rightarrow 0$.
Therefore in the infrared we are left with only the gauge field
degrees of freedom and the theory should be effectively
the same as pure QCD$_3$.

According to Maldacena \cite{mal}, the
${\cal N}=4$ theory in Euclidean ${\bf R}^4$
is dual to type IIB superstring theory on AdS$_5 \times {\bf S}^5$
with the metric
\beq
 {ds^2 \over l_s^2 \sqrt{4 \pi g_s N}} = \rho^{-2} d\rho^2 +
\rho^2
\sum_{i=1}^4 dx_i^2 + d \Omega_5^2
\label{ads5}
\eeq
where $l_s$ is the string length related to the superstring
tension, $g_s$ is the string coupling constant
and $d\Omega_5$ is the line element on ${\bf S}^5$. The
$x^{1,2,3,4}$ directions in AdS$_5$ correspond  to ${\bf R}^4$
where the gauge theory lives. The gauge coupling constant $g_4$ of
the 4D theory is related to the string
coupling constant $g_s$ as $g_4^2 = g_s$.
In the 't Hooft limit
($N \rightarrow \infty$ with $g_4^2N = g_s N$ fixed),
the string coupling constant vanishes $g_s \rightarrow 0$.
Therefore we can study the 4D theory using the
first quantized string theory in the AdS space
(\ref{ads5}). Moreover if $g_s N \gg 1$, the curvature
of the AdS space is small and the string theory is approximated
by classical supergravity.

Upon compactification on ${\bf S}^1$ and  imposing
the supersymmetry breaking boundary conditions, (\ref{ads5})
is replaced by the Euclidean black hole geometry \cite{w2}
\beq
  {ds^2 \over l_s^2 \sqrt{4 \pi g_s N}}
=
\left(\rho^2 - {b^4 \over \rho^2}
\right)^{-1} d\rho^2 +
\left(\rho^2 - {b^4 \over \rho^2}
\right) d \tau^2 + \rho^2
\sum_{i=1}^3 dx_i^2 + d\Omega_5^2
\label{metric}
\eeq
where $\tau$ parameterizes the compactifying circle and
the $x^{1,2,3}$ direction corresponding to the ${\bf R}^3$
where QCD$_3$ lives. The horizon of this geometry is
located at $\rho=b$ with
\beq
    b = {1 \over 2R}.
\label{horizonlocation}
\eeq
Once again, the supergravity approximation is applicable
for $N \rightarrow \infty$ and $g_s N \gg 1$.

According to \cite{GKP,witten-one,ho,f,ozterning},
there is a one-to-one correspondence between supergravity wave solutions
on AdS$_5 \times {\bf S}^5$ and chiral primary fields
(and their descendants) in the ${\cal N}=4$ superconformal
theory in four dimensions.
The mass $m$  of a $p$-form $C$ on the AdS space
is related to the dimension $\Delta$ of a $(4-p)$ form
operator in the ${\cal N}=4$ theory by
\beq
 m^2 = (\Delta-p)(\Delta+p-4) .
\label{dimp}
\eeq

The supergravity fields on AdS$_5 \times {\bf S}^5$ can be
classified by decomposing them into spherical harmonics
(the Kaluza-Klein modes) on ${\bf S}^5$. They fall into
irreducible representations
of the $SO(6)$ isometry group of ${\bf S}^5$,
which
is also the $R$-symmetry group of
the 4D superconformal theory.
The spectrum of Kaluza-Klein harmonics of type IIB supergravity
on AdS$_5\times {\bf S}^5$ was derived in \cite{van0,gm}. Among them,
there are four Kaluza-Klein
modes that are $SO(6)$ singlets, coming from the $s$-wave
components on ${\bf S}^5$ of bosonic fields. They are:

\noindent
(1) The graviton $g_{\mu\nu}$ polarized
along the ${\bf R}^4$ in (\ref{metric}). It couples
to the dimension 4 stress-energy tensor $T_{\mu\nu}$
of the $\cN=4$  theory.

\noindent
(2) The dilaton and the R-R scalar, which combine into
a complex massless scalar field. Its real and imaginary parts
couple to the dimension 4 scalar operators $\cO_4 = {\rm tr}~ F^2$ and
$\tilde \cO_4 = {\rm tr} ~F\wedge F$
of the  $\cN=4$ theory respectively.

\noindent
(3) The NS-NS and R-R two-forms, which combine into a complex-valued
 antisymmetric field $A_{\mu\nu}$,
polarized along the ${\bf R}^4$. Its $({\rm AdS~ mass})^2=16$ and
using (\ref{dimp}) we see that it couples to a dimension 6 two-form
operator of the $\cN=4$ theory. This operator has been identified as
$\cO_6 = d^{abc} F^a_{\mu \alpha} F^{b\alpha \beta} F^c_{\beta \nu}$
\cite{dastrivedi,flz}.

\noindent
(4) The $s$-wave component of the metric $g^\alpha_\alpha$ and
the R-R 4-form $A_{\alpha\beta\gamma\delta}$ polarized
along ${\bf S}^5$. They combine into a massive scalar with
$({\rm AdS~ mass})^2=32$ and couple
to a dimension 8 scalar operator constructed from the gauge
field strength $F_{\mu\nu}$ of the $\cN=4$ theory \cite{aki,flz}.

\noindent
Only these $SO(6)$ singlet fields are related to glueballs
of QCD$_3$ since $SO(6)$ non-singlets are supposed to decouple
in the limit $R \rightarrow 0$.

Let us discuss now how to identify the quantum numbers of the glueballs.
The spin and the parity of a glueball in three dimensions
can be easily found from the transformation
properties of the corresponding supergravity field.
The charge conjugation, $C$, for gluons
is defined by $A_{\mu}^aT^a_{ij} \rightarrow -A_{\mu}^a
T^a_{ji}$ where the $T^a$'s are
the hermitian generators of the gauge group \cite{mandula}.
In the string theory, charge conjugation corresponds
to the worldsheet parity transformation changing the orientation
of the open string attached to D-branes.
Therefore, for example, the NS-NS two-form in supergravity
is odd under the charge conjugation. This is consistent with the fact that
it couples to $\cO_6$, which indeed has $C=-1$.

 From the point of view of QCD$_3$, the radius $R$ of the
compactifying circle provides  the ultraviolet cutoff
scale. To obtain large $N$
QCD$_3$ in the continuum, one has to take $g_4^2 N \rightarrow 0$
as $R \rightarrow 0$ so that $g_3^2 N = g_4^2N/R$ remains
at the intrinsic energy scale of QCD$_3$.
Here $g_3$ is the dimensionful gauge coupling of  QCD$_3$.
This is the opposite of the limit
that is required for the supergravity description to be valid.
As we mentioned, the supergravity description is applicable
for $g_4^2 N \gg 1$.
Therefore, with the currently available techniques,
the Maldacena-Witten conjecture
can only be used to study large $N$ QCD with a fixed ultraviolet
cutoff $R^{-1}$ in the strong ultraviolet coupling regime.
The results we find are, however, surprisingly close to those of
the lattice computation, leading us to suspect that $(g_4^2N)^{-1}$
corrections to the mass ratios are small. In section 4, we will estimate
the leading $(g_4^2N)^{-1}$ correction to our computation.

Consider first the  $0^{++}$ glueball masses. These can be derived from the
2-point function of the operator ${\rm tr}~ F_{\mu \nu} F^{\mu\nu}$.
In the supergravity description we have to solve the classical
equation of motion of the massless dilaton,
\beq
\label{dilaton}
\partial_{\mu} \left[ \sqrt{g} \partial_{\nu}\Phi
g^{\mu \nu} \right] =0 \comma
\eeq
on the AdS$_5$ black hole background (\ref{metric}).
In order to find the lowest mass modes we assume following \cite{w2}
that $\Phi$ is independent
of $\tau$ and has the form $\Phi =f(\rho )e^{ikx}$. Using the metric of
(\ref{metric}) one obtains the following differential equation for
$f$:
\beq
\rho^{-1} {{d}\over{d \rho}} \left( \left(\rho^4 -
b^4 \right) \rho {{d f}\over{d \rho}} \right) - k^2 f = 0
\stop
\label{dilatondiff}
\eeq
Since the glueball mass $M^2$ is equal to $-k^2$, the task
is to solve this equation as an eigenvalue problem for $k^2$.
In the following we set $b=1$, so the masses are computed
in units of $b$.
If one changes variables to  $x = \rho^2$,
the equation takes the form,
\beq
  {d^2 f \over dx^2} + \left( {1 \over x} +
{1 \over x-1} + {1 \over x+1} \right) {d f \over dx}
- {k^2 \over 4x(x^2-1)} f = 0,
\label{regularsing}
\eeq
namely it is an ordinary differential equation with four regular
singularities at $x=0, \pm 1$ and $\infty$.

Unlike the equation with three regular singularities (known as
the hypergeometric equation), analytic solutions are not
known for this type of equation.
Fortunately there is an analytical method to compute its
eigenvalues $k^2$.  It is the exact WKB analysis recently developed
by mathematicians at RIMS, Kyoto University \cite{takei}.
To use their approach, we note that the differential equation
(\ref{regularsing}) can be written as the Schr\"odinger-type
equation
\beq
   \left( - {d^2 \over dx^2} + Q(x) \right) g(x) = 0,
\eeq
where $g(x) = \sqrt{x(x^2-1)} f(x)$ and
\beq
  Q(x) =
   {3x^4 - 6x^2 -1 \over 4x^2(x^2-1)^2} +
    {k^2 \over 4x(x^2-1)} .
\eeq
To apply the WKB analysis, one can perturb the equation
as
\beq
   \left(  - {d^2 \over dx^2} + Q(x) +
(\eta^2 - 1) R(x) \right) g(x) = 0,
\eeq
by introducing a large parameter $\eta$.
With a suitable choice of $R(x)$, the secular equation,
which determines the values of $k^2$ so that the
equation admits a solution regular at both $x=1$ and $\infty$,
becomes explicitly solvable as a asymptotic power series expansion
in $\eta^{-1}$.
Assuming the expansion is Borel summable at $\eta=1$,
the eigenvalues are approximated by the following
expression \cite{tk}
\beq
k^2 = - 6 n(n+1) ~,~~~(n=1,2,3,...).
\label{wkb}
\eeq
We should note that the differential
equation in question is degenerate from the point of view
of the exact WKB analysis and a mathematical proof
of the Borel summability in this case
has not been given. It is possible that
the formula (\ref{wkb}) receives small corrections.

Since the analytical expression (\ref{wkb}) for $k^2$ is
still preliminary and we would like to find masses
for the other glueball states, we also solved the differential
equation (\ref{dilatondiff}) numerically.
For large $\rho$, the black hole
metric (\ref{metric}) asymptotically approaches
the AdS metric, and the
behavior of the solution for a $p$-form
for large $\rho$ takes the form  $\rho^{\lambda}$, where $\lambda$ is
determined from the mass $m$ of the supergravity field:
\beq
  m^2 = \lambda( \lambda + 4 - 2p)~.
\label{power}
\eeq
Indeed both (\ref{dilatondiff}) and (\ref{power}) give the
asymptotic forms $f \sim 1, \rho^{-4}$, and
only the later is a normalizable solution~\cite{w2}.
Changing variables to $f=\psi/\rho^4$ we have:
\beq
 \left( {\rho^2} - {\rho^6} \right) \,\psi^{\prime\prime}    +
  \left( 3\,{\rho^5} -7 \rho  \right) \,\psi^\prime
+\left( 16  + k^2  \rho^2 \right) \, \psi = 0
\stop
\eeq
For large $\rho$ this equation can be solved by series
solution with negative even powers:
\beq
\psi = \Sigma_{n=0}^\infty a_{2n} \rho^{-2n}
\label{asymp}
\stop
\eeq
Since the normalization is arbitrary we can set $a_0=1$.  The
first few coefficients are given by:
\begin{eqnarray}
a_2 &=& {{k^2}\over{12}} \nonumber\\
a_4 &=& {{1}\over{2}} + {{k^4}\over{384}} \nonumber\\
a_6 &=& {{7 k^2}\over{120}} + {{k^6}\over{23040}}
\stop
\label{asymp2}
\end{eqnarray}
For $n \ge 5$ the coefficients are given by the recursive relation:
\beq
(n^2+4n) a_n = k^2 a_{n-2} + n^2 a_{n-4} ~.
\eeq
Since the black hole geometry is regular at the horizon
$\rho=1$, $k^2$ has to be adjusted so that
$f$ is also regular at $\rho=1$ \cite{w2}\footnote{We thank A. Jevicki and
J. P. Nunes for communication on the boundary condition.}.  
This can be done numerically in a simple fashion using 
a ``shooting" technique as follows.
For a given value of $k^2$ the equation is numerically integrated from
some sufficiently
large value of $\rho$ ($\rho \gg k^2$) by matching
$f(\rho)$ with the asymptotic
solution set by (\ref{asymp}) and (\ref{asymp2}).
The glueball mass $M$ is related to
the eigenvalues of $k^2$ by $M^2 = - k^2$ in units of $b^2$.
The results of the numerical work are listed in Table 1.
They agree with the formula (\ref{wkb}). The 4$\%$ discrepancy
of the two results are either due to some systematic error
in the numerical analysis or due to corrections to the analytical
formula (\ref{wkb}).

\begin{table}[htbp]
\centering
\begin{tabular}{l|ccc}
state &  numerical method & exact WKB method &
ratio \\ \hline
$0^{++}$ & 11.59 & 12 & 1.03 \\
 $0^{++*}$ & 34.53 & 36 & 1.04 \\
 $0^{++**}$ & 68.98 & 72 & 1.04 \\
 $0^{++***}$ & 114.9 & 120 & 1.04 \\
 $0^{++****}$ & 172.3 & 180 & 1.04 \\
 $0^{++*****}$ & 241.2 & 252 & 1.04 \\
\end{tabular}
\label{summary1}
\parbox{4in}{\caption{(Mass)$^2$ of $0^{++}$ glueball in QCD$_3$
obtained by solving the supergravity wave equation
in the black hole geometry (in units of $b^2$) using
the two different methods.}}
\end{table}

Since both methods give the same results within a 4$\%$ error,
we are ready to compare them with the lattice gauge theory
computations \cite{teper}. Since the lattice results are
in units of string tension, we normalize the supergravity
results  so that the lightest $0^{++}$ state agrees with
the lattice result. The results are listed in Table 2.
 One should also expect a systematic error in addition to the  statistical
error
denoted in Table 2 for the lattice computations.

\begin{table}[htbp]
\centering
\begin{tabular}{l|ccc}
state & lattice, $N=3$ & lattice, $N\rightarrow \infty$ &
supergravity  \\
 \hline
 $0^{++}$ & $4.329 \pm 0.041$ & $4.065 \pm 0.055$ & 4.07 ({\rm input}) \\
 $0^{++*}$ & $6.52 \pm 0.09$ & $6.18 \pm 0.13$ & 7.02 \\
 $0^{++**}$ & $8.23 \pm 0.17$ & $7.99 \pm 0.22$ & 9.92 \\
 $0^{++***}$ &  - & - & 12.80 \\
 $0^{++****}$ &  - & - & 15.67 \\
 $0^{++*****}$ & -  & - & 18.54 \\
\end{tabular}
\parbox{4in}{\caption{$0^{++}$ glueball masses in QCD$_3$
coupled to ${\rm tr}~F_{\mu \nu}
F^{\mu\nu}$. The lattice results are in units of the square root
of the string tension. The denoted
error in the lattice results is only the statistical one.}\label{summary}}
\end{table}

Next we consider the two-form of the supergravity theory.
As noted previously, it couples to the operator $\cO_6$.
This operator contains $1^{+-}$ and $1^{--}$ components,
which correspond to the fields $A_{\tau i}$ and $A_{ij}$, where
$i,j=1,2,3$ correspond to the three coordinates $x_i$
of ${\bf R}^3$.
The remaining components $A_{\rho \tau}$ and $A_{\rho i}$ can be set
to zero by an appropriate gauge transformation.
In the QCD$_3$ limit $R \rightarrow 0$,
the  $1^{--}$ component $A_{ij}$
is reduced to a $0^{--}$ operator
in 3D, and thus has a non-zero overlap
with the  $0^{--}$ glueball\footnote{The parity $P=-1$
is due to the fact that
the 2-form is dual to a pseudoscalar. The charge conjugation
$C=-1$ is inferred from the string worldsheet parity.}.
On the other hand,
the $1^{+-}$ components $A_{\tau i}$ couple to an operator
which is supposed to decouple in the
$R  \rightarrow 0$ limit. Therefore they do not
correspond to glueball states in QCD$_3$.

The $s$-wave component of the two-form field satisfies the equation
\cite{van0,mathur}
\beq
{{3}\over{\sqrt{g}}} \partial_\mu\left[\sqrt{g} \, \partial_{[\mu^\prime}
A_{\mu_1^\prime
\mu_2^\prime]} \, g^{\mu^\prime \mu} g^{\mu_1^\prime \mu_1} g^{\mu_2^\prime
\mu_2}\right]
- 16  g^{\mu_1^\prime \mu_1} g^{\mu_2^\prime \mu_2}
A_{\mu_1^\prime \mu_2^\prime} = 0 \comma
\eeq
where $[~~]$ denotes antisymmetrization with strength one.
As before we  look for solutions which are independent of $\tau$
and are of the form $A_{ij}= h_{ij}(\rho )e^{ikx}$.
The $\rho \tau$ and the $\rho i$ components of this equation
simply result in a constraint which sets
the transverse component of $A_{ij}$ to zero.
For the remaining
pseudoscalar component from $A_{ij}$ the equation reduces to
\beq
 \rho\left({\rho^4}\, -  1\right) h''
 + \left( 3 + {\rho^4}\right)h'  -\left( {k^2}\,\rho\,
+16\,{\rho^3} \right)  h =0 ~,
\eeq
in units where $b=1$.
At large $\rho$ the solution has the form
\beq
h = \rho^{-4} \, \Sigma_{n=0}^\infty a_{2n} \rho^{-2n}
\stop
\label{pseudoasymp}
\eeq
Since the normalization is arbitrary we can again set $a_0=1$.  The
first few coefficients are given by:
\begin{eqnarray}
a_2 &=& {{k^2}\over{20}} \nonumber\\
a_4 &=&  {{640 +k^4}\over{960}} \nonumber\\
a_6 &=& {{3520 k^2 + k^6}\over{80640}}
\stop
\end{eqnarray}
We have solved the differential equation and obtained
the eigenvalues $k^2$ by the same numerical
method described above. The
results are shown in Table 3. The supergravity
results are displayed in the same
normalization as the one used in Table 2.

\begin{table}[htbp]
\centering
\begin{tabular}{l|ccc}
  state & lattice, $N=3$ & lattice, $N\rightarrow \infty$ &
supergravity  \\
 \hline
 $0^{--}$ &$6.48 \pm 0.09$ &$5.91 \pm 0.25$ & 6.10 \\
 $0^{--*}$ &$8.15 \pm 0.16$ & $7.63 \pm 0.37$ & 9.34 \\
 $0^{--**}$ & $9.81 \pm 0.26$ & $8.96 \pm 0.65$& 12.37 \\
 $0^{--***}$ & - & - &  15.33 \\
 $0^{--****}$ & - & -  & 18.26 \\
 $0^{--*****}$ & - &  - & 21.16 \\
\end{tabular}
\label{3dtensor}
\parbox{4in}{\caption{
{$0^{--}$ glueball masses in QCD$_3$
coupled to $\cO_6$.
The lattice results are in units of square root of the string
tension. The normalization of the supergravity results
is the same as in Table 2.}}}
\end{table}

Since the supergravity method and the lattice gauge theory
compute the glueball masses in different units,
one cannot compare the absolute values of the
lowest glueball mass obtained using these methods.
However it makes sense to compare the lowest glueball
masses of different quantum numbers.
Using Tables 2 and 3, we find that the supergravity
results
are in good agreement with the lattice gauge theory
computation \cite{teper}:
\beqar
&\left(\frac{M_{0^{--}}}{M_{0^{++}}}\right)_{{\rm supergravity}}&= 1.50 \non\\
&\left(\frac{M_{0^{--}}}{M_{0^{++}}}\right)_{{\rm lattice~~~~~}}& =
 1.45\pm 0.03
\stop
\eeqar

There are still two more $SO(6)$ singlet supergravity fields.
One of them is the $s$-wave component of the metric $g^\alpha_\alpha$
and the R-R 4-form $A_{\alpha\beta\gamma\delta}$
polarized along ${\bf S}^5$. From (\ref{dimp}) we see that
it should couple to a dimension 8 scalar operator $\cO_8$. In
\cite{flz,aki}, this operator is identified as
a symmetrized form of
$\left[F^4 - {1 \over 4} (F^2)^2\right]$. By using
the prescription of Tseytlin \cite{ts} to symmetrize
the group indices, one finds that the operator
is even under the charge conjugation. This
is also seen from the fact that  $g^\alpha_\alpha$
is clearly even both spacetime and worldsheet parity
transformations. Therefore $g^\alpha_\alpha$ has the quantum
numbers of the $0^{++}$ glueball.
The classical equation of motion of  $g^\alpha_\alpha$ is that of
a massive scalar with (AdS mass)$^2 = 32$
(in units of $b^2$) \cite{van0} on the AdS$_5$ black
hole background (\ref{metric}).
The mass spectrum that we get is given in Table 4.
In the  $g_s N \rightarrow \infty$ limit the operators ${\cal O}_8$ and
tr$F^2$ are not mixed since they couple to different states in the
supergravity theory.  However, we expect that for finite $g_sN$ these
operators will mix, thus the full $0^{++}$ spectrum is expected to be
given by the interleaving of Tables \ref{summary} and \ref{3d4tensor}.
For example the $0^{++**}$ presumably corresponds to the first state in
Table \ref{3d4tensor}.

\begin{table}[htbp]
\centering
\begin{tabular}{c}
 $g^{\alpha}_\alpha$ and  $A_{\alpha\beta\gamma\delta}$ \\
\hline
 8.85\\
12.06\\
 15.00\\
 17.98\\
\end{tabular}

\parbox{4in}{\caption{$0^{++}$ glueball masses in QCD$_3$
coupled to $\cO_8$, the normalization is the same as in Table
\protect\ref{summary}.}\label{3d4tensor}}
\end{table}

The remaining $SO(6)$ singlet is the graviton $g_{\mu\nu}$.
It couples to the energy-momentum tensor $T_{\mu\nu}$ and
therefore corresponds
to the $2^{++}$ glueball. It would be interesting
to compute its mass and compare with the lattice result.

\section{Glueballs in Four Dimensions}

To construct QCD$_4$, one starts with
the superconformal theory in six dimensions realized
on $N$ parallel coinciding M$5$-branes. The compactification of this theory on
a circle of radius $R_1$ gives
a five-dimensional theory whose low-energy effective
theory is the maximally supersymmetric $SU(N)$ gauge
theory with gauge coupling constant $g_5^2 = R_1$.
To obtain QCD$_4$, one compactifies this theory further on
another ${\bf S}^1$ of radius $R_2$. The gauge coupling
constant $g_4$ in 4D is given by
$g_4^2 = g_5^2/R_2 = R_1/R_2$. To break supersymmetry,
one imposes the anti-periodic boundary condition on
the fermions around the second ${\bf S}^1$.

According to Maldacena \cite{mal}, the large $N$ limit of the six-dimensional
theory is $M$ theory on AdS$_7 \times {\bf S}^4$. Upon
compactification on ${\bf S}^1 \times {\bf S}^1$ and imposing
the anti-periodic boundary conditions around the second ${\bf S}^1$,
we find $M$ theory to be on the black hole geometry \cite{w2}.
To take the large $N$ limit while keeping $g_4^2 N$ finite,
we have to take $R_1 \ll R_2$. In this limit, $M$ theory
reduces to type IIA string theory and the M$5$ brane wrapping
on ${\bf S}^1$ of radius $R_1$ becomes a D$4$ brane. The large
$N$ limit of QCD$_4$ then becomes string theory on the
black hole geometry given by
\beq
{ds^2 \over l_s^2 g_5^2 N/4\pi} = {d \rho^2 \over 4 \rho^{3/2}
 \left(1 - {b^6 \over \rho^3}\right) }
+ \rho^{3/2} \left(1 - {b^6 \over \rho^3}
\right) d \tau^2 + \rho^{3/2}
\sum_{i=1}^4 dx_i^2 +  \rho^{1/2} d\Omega_4^2,
\label{D4}
\eeq
with a dilaton $e^\phi \sim \rho^{3/4}$ \cite{imsy}. The location of the
horizon $\rho=b^2$ is related to the radius $R_2$ of
the compactifying circle as
\beq
    b = {1 \over 3 R_2}
\eeq
As in the case of three dimensions,
 we will compute the spectrum of glueball masses by solving the
classical equations of motions of Kaluza-Klein modes of the supergravity
theory.
We will consider only singlets of the $SO(5)$ isometry group
of ${\bf S}^4$, which corresponds to the $R$-symmetry group of
the six-dimensional theory.

Consider first the
$0^{++}$ glueball. The non-extremal D$4$ brane solution has a non constant
dilaton background.
As shown in \cite{hs}
the dilaton is a linear combination of two scalars. One of
them is massless and couples to the relevant glueball operator.
The equation of motion for the scalar  is
given by (\ref{dilaton})  in the background of the
metric (\ref{D4}). Again assuming that the  solution
is independent of $\theta$ and of the form
$\Phi =f(\lambda)e^{ikx}$ (with $\lambda^2 = \rho$),
one obtains the differential equation in the units where  $b=1$ as
\beq
(\lambda^7-\lambda)f''(\lambda )+(10\lambda^6-4)f'(\lambda )-\lambda^3k^2
f(\lambda )=0
\stop
\eeq
The asymptotic solutions to this equation are $f \sim 1,
\lambda^{-9}$, with the latter corresponding to normalizable solutions.
In order to solve the equation and find the
allowed values of $k^2$ we introduce the function $g (\lambda )$ as
$f(\lambda )=\lambda^{-9} g(\lambda )$.
This way $g(\lambda)$ has to be asymptotically
constant for $\lambda \rightarrow \infty$,
and one can again look for a solution in terms of a negative power
series in $\lambda$. The differential equation for $g(\lambda)$ is
\beq
\label{geq}
(\lambda^8-\lambda^2)g''+(14\lambda -8\lambda^7)g'-(\lambda^4k^2+54)g=0.
\eeq
The first few coefficients in the power series solution
$g=\sum_{n=0}^\infty a_{2n} \lambda^{-2n}$ are given by (for $a_0=1$)
\begin{eqnarray}
a_2 &=& {{k^2}\over{22}} \nonumber\\
a_4 &=&  {{k^4}\over{1144}} \nonumber\\
a_6 &=& \frac{61776+k^6}{102960}
\stop
\end{eqnarray}
The regularity of $f$ at $\lambda=1$, after numerically solving the
equation (\ref{geq}) as described in the previous section, results
in the allowed values of $k^2$. The first six masses (normalized so that the
lightest $0^{++}$ state agrees with the lattice calculation)
together with the available lattice results \cite{teper4,morning} are given in
Table 5.

\begin{table}[htbp]
\centering
\begin{tabular}{l|cc}
state & lattice, $N=3$ &
supergravity  \\
 \hline
 $0^{++}$ & $1.61 \pm 0.15$   & 1.61 {\rm (input)} \\
 $0^{++*}$ &  2.8   & 2.38 \\
 $0^{++**}$ &   - & 3.11 \\
 $0^{++***}$ &  -  & 3.82 \\
 $0^{++****}$ &  -  & 4.52 \\
 $0^{++*****}$ &  -  & 5.21 \\
\end{tabular}
\parbox{4in}{\caption{Masses of the first few $0^{++}$ glueballs in
QCD$_4$, in GeV,
from supergravity compared
to the available lattice results. Note that the authors
of
ref. \protect\cite{morning} do not quote errors for the $0^{++*}$ since it is
not yet clear whether
it is a genuine excited state or merely a two glueball bound state.}}
\label{tab:4ddila}
\end{table}

In order to calculate the masses of the $0^{-+}$ glueball in four dimensions
 we will consider  the
3-form $A_{\alpha\beta\gamma}$ of the eleven dimensional supergravity.
In this case, it is more useful to use the eleven-dimensional metric
\beq
  ds^2 = {d \lambda^2 \over
 \left( {\lambda^2 \over b^2} - {b^4 \over \lambda^4} \right)} +
\left( {\lambda^2 \over b^2} - {b^4 \over \lambda^4} \right) d\tau^2
 + \lambda^2 \sum_{i=1}^5 dx_i^2 + d \Omega_4^2,
\eeq
which reduces to (\ref{D4}) upon compactifying $x_5$ on
${\bf S}^1$ and by going to the string frame \cite{w2}
by multiplying the metric by $\lambda$, setting $\lambda^2 =\rho$, and
rescaling
the other coordinates.
The $s$-wave component of the 3-form in the harmonic
expansion
on ${\bf S}^4$ is a singlet of the $SO(5)$ isometry group \cite{van1}.
Its mass squared\footnote{The value of the mass term
is fixed by matching the supergravity computation \cite{van1} to
(\ref{dimp}) \cite{AOY}.} is
$36$ in units of $b^2$ and using (\ref{dimp}) we see that it
couples to a dimension 9 operator of the six-dimensional theory.
The 3-form obeys the
following equation of motion:
\beq
{{4}\over{\sqrt{g}}} \partial_\mu\left[\sqrt{g} \, \partial_{[\mu^\prime}
A_{\mu_1^\prime
\mu_2^\prime \mu_3^\prime]}
\, g^{\mu^\prime \mu} g^{\mu_1^\prime \mu_1} g^{\mu_2^\prime
\mu_2}g^{\mu_3^\prime \mu_3}\right]
-36  g^{\mu_1^\prime \mu_1} g^{\mu_2^\prime \mu_2}
g^{\mu_3^\prime \mu_3}
A_{\mu_1^\prime \mu_2^\prime \mu_3^\prime} = 0
\stop
\eeq
Choosing a gauge where $A_{\rho \tau i}$ and $A_{\rho ij}$ vanish,
where $i,j=1,\ldots ,5$, and assuming that the remaining components
are independent of the coordinate $\tau$ and the only dependence
on $x_i$ is through $e^{ikx}$, one finds that there are
two independent modes after compactification to 4D:

\noindent
(1)  A three-index tensor
$A_{ijk}$. This is dual to
a massive scalar and can be identified with the
$0^{-+}$ glueball of the 4D theory.

\noindent
(2) A  massive vector
$A_{\tau ij}$. This couples to an operator which
is supposed to decouple in the limit $R_2 \rightarrow 0$.
Therefore it does not correspond to a glueball state in
QCD$_4$.

The scalar component of $A_{ijk}$ satisfies the differential equation
\beq
(\lambda^7-\lambda )f''(\lambda )
+(\lambda^6+5)f'(\lambda )-\lambda^3(k^2+36\lambda^2)f(\lambda )=0,
\eeq
in the same units as in the equation (3.3) for the dilaton.
The normalizable asymptotic solution behaves like $1/\lambda^6$, thus we
introduce the function $g(\lambda )$ by $f(\lambda ) = \lambda^{-6}
g(\lambda)$.
This satisfies
\beq
\label{geq2}
(\lambda^7-\lambda)g''(\lambda )-(11\lambda^6-17)g'(\lambda )
-(72+k^2\lambda^4)g(\lambda )=0.
\eeq
The power series expansion $g(\lambda )=\sum_{n=0}^\infty a_{2n}
\lambda^{-2n}$
with $a_0=1$ has the first few coefficients
\begin{eqnarray}
a_2 &=& {{k^2}\over{28}} \nonumber\\
a_4 &=&  {{k^4}\over{1792}} \nonumber\\
a_6 &=& \frac{129024+k^6}{193536}
\stop
\end{eqnarray}
The single-valuedness of the solution requires
$g' = 6g$ at $\lambda=1$. With this boundary condition,
the numerical solution of (\ref{geq2})
yields the allowed values
of $k^2$, and the corresponding masses in the units
defined above
are displayed in Table 6.
\begin{table}[htbp]
\centering
\begin{tabular}{l|cc}
  state & lattice, $N=3$  &
supergravity  \\
 \hline
 $0^{-+}$ &  $2.19 \pm 0.32$& 1.83 \\
 $0^{-+*}$ &  - & 2.67 \\
 $0^{-+**}$ &  - & 3.42 \\
 $0^{-+***}$ & - & 4.14 \\
 $0^{-+****}$ &  -  & 4.85 \\
 $0^{-+*****}$ & -  & 5.55 \\
\end{tabular}
\label{4dtensor}
\parbox{4in}{\caption{Masses of  $0^{-+}$ glueball in QCD$_4$. The lattice
result is in GeV.}}
\end{table}

Unlike the 3D case, there exists little lattice data on the masses of the
excited glueball states.
We can however compare the ratio of masses of the lowest glueball states
$0^{-+}$ and $0^{++}$
\beqar
&\left(\frac{M_{0^{-+}}}{M_{0^{++}}}\right)_{{\rm supergravity}}&= 1.14 \non\\
&\left(\frac{M_{0^{-+}}}{M_{0^{++}}}\right)_{{\rm lattice~~~~~}}& =
1.36 \pm 0.32
\comma
\eeqar
and the results are in agreement within the one $\sigma$ error.

\section{Leading String Theory Corrections}

As we mentioned earlier, the supergravity computation is
valid in the strong ultraviolet coupling limit $g_s N \gg 1$. In order to
compare with the lattice computations in the continuum limit,
we have to take $g_s N \rightarrow 0$ as we take the ultraviolet
cutoff $R^{-1} \rightarrow \infty$
so that the scale set by
the Yang-Mills coupling constant remains at the intrinsic energy
scale of QCD. The fact that the glueball masses computed in the
supergravity limit are in good agreement with the lattice results
leads us to suspect that, for this particular computation,
$\alpha'$ corrections are small. In this section, we test
this idea.

For $g_s N \ll 1$, the curvature of the black hole
geometry becomes larger than the string scale. Therefore
stringy corrections (to be precise, the worldsheet sigma-model
corrections) are expected to become important. The leading
stringy corrections to the low-energy supergravity action were
obtained in \cite{grisaru,gw}.
Recently Gubser, Klebanov and Tseytlin \cite{GKT} used
the modified action to obtain the leading order string corrections
to the black hole metric.
We use their result to calculate the leading corrections to
the glueball mass spectrum. We will perform this computation
only for the $0^{++}$ glueballs in QCD$_3$. We expect, however,
that the conclusions will be similar for the other glueball states.

According to \cite{GKT}, the leading (in units of the curvature)
$\alpha' = (4\pi g_s N)^{-1/2}$
correction to the AdS$_5$ black hole metric (\ref{metric})
is
\beq
  {ds^2 \over l_s^2 \sqrt{4 \pi g_4^2 N}}
=
(1+\delta_1) {d\rho^2 \over \left(\rho^2 - {b^4 \over \rho^2}
\right)}+ (1+\delta_2)
\left(\rho^2 - {b^4 \over \rho^2}
\right) d \tau^2 + \rho^2
\sum_{i=1}^3 dx_i^2,
\label{correction}
\eeq
where the correction terms $\delta_{1,2}$ are given by the formulae
\begin{eqnarray}
\delta_1& = & - 15\gamma \left( 5\frac{b^4}{\rho^4}+5\frac{b^8}{\rho^8}-
3\frac{b^{12}}{\rho^{12}}\right) \nonumber \\
\delta_2& = & + 15\gamma \left( 5\frac{b^4}{\rho^4}+5\frac{b^8}{\rho^8}-
19\frac{b^{12}}{\rho^{12}}\right),
\end{eqnarray}
and $\gamma$ is given by $\gamma =\frac{1}{8}\zeta (3) \alpha'^3$.
With these corrections of the metric, the dilaton
is no longer constant, instead it is given by
\beq
\Phi_0 = -\frac{45}{8}\gamma \left( \frac{b^4}{\rho^4}+\frac{b^8}{2\rho^8}+
\frac{b^{12}}{3\rho^{12}}\right).
\label{dilbackgr}
\eeq

There is also a correction to the ten-dimensional dilaton action
\cite{grisaru,gw}, given
by
\beq
I_{dilaton}=-\frac{1}{16\pi G_{10}} \int d^{10}x
\sqrt{g} \left[ -\frac{1}{2}g^{\mu\nu}
\partial_{\mu}\Phi \partial_{\nu}\Phi
+\gamma e^{-\frac{3}{2}\Phi} W \right],
\label{action}
\eeq
where $W$ is given in terms of the Weyl tensor, and in our
background $W = 180/\rho^{16}$ in units where  $b=1$. To
the leading order in $\gamma$, the dilaton perturbation
does not mix with the metric perturbation, so we can
study the dilaton equation derived from the action (\ref{action})
in the fixed metric background (\ref{correction}). In subleading
order in $\gamma$,
the term $\gamma e^{-\frac{3}{2}\Phi} W$ would generate
a mixing of the dilaton and the graviton.

We now have all pieces needed in order to obtain the
first order correction to the dilaton equation.
We write $\Phi = \Phi_0 +f(\rho)e^{ikx}$, with $\Phi_0$ given by
(\ref{dilbackgr}), and expand $f(\rho)$ and $k^2$ in $\gamma$ as
\beq
f(\rho )=f_0(\rho )+\gamma h(\rho )~,~~
k^2=k_0^2+\gamma \delta k^2.
\eeq
Here $f_0 (\rho )$ obeys the lowest order equation (\ref{dilatondiff})
and is a numerically given function, and $k_0^2$ are the eigenvalues
numerically obtained from the solution of (\ref{dilatondiff}).
The first order term of the
differential equation obtained from the action (\ref{action})
using the background (\ref{correction}) and (\ref{dilbackgr})
is given by
\begin{eqnarray}
 \rho^{-1} {d \over d\rho} \left(
(\rho^4 - 1) \rho {d h \over d\rho} \right)
- k_0^2 h &=&
(75 - 240 \rho^{-8}  + 165 \rho^{-12})
{d^2 f_0 \over d\rho^2}  \nonumber \\
&& + (75 + 1680 \rho^{-8} - 1815 \rho^{-12}) \rho^{-1}
 {d f_0 \over d\rho} \nonumber \\
& &+ ( \delta k^2   - 120   k^2_0
\rho^{-12}  - 405 \rho^{-14}  ) f_0(\rho).
\label{inhomo}
\end{eqnarray}
With $f_0(\rho)$ and $k_0^2$ given, one may regard
this as an inhomogeneous version of the equation
(\ref{dilatondiff}).
We solve this equation to determine $h(\rho)$
and $\delta k^2$.

At large $\rho$ the solution for the first order correction has the form
\beq
h = \rho^{-4} \, \Sigma_{n=0}^\infty b_{2n} \rho^{-2n}
\stop
\label{pseudoasymp2}
\eeq
Since (\ref{inhomo}) is inhomogeneous for $h(\rho)$,
one can always add to a solution $h(\rho)$
a constant multiple of the solution
$f_0(\rho)$ to the corresponding homogeneous equation (\ref{dilatondiff})
to obtain another solution. We use this freedom
to set $b_0=0$.  The first few coefficients are then given by:
\begin{eqnarray}
b_2 &=& {{\delta k^2}\over{12}} \nonumber\\
b_4 &=&  {{14400 +2 \delta k^2 k_0^2}\over{384}} \nonumber\\
b_6 &=& {{1344 \delta k^2 +100800 k_0^2 +3 \delta k^2 k_0^4}\over{23040}}
\stop
\end{eqnarray}
We can now determine
$\delta k^2$ by the same ``shooting" method described above
for each
eigenvalue of $k_0^2$ and its corresponding eigenfunction $f_0(\rho)$.
It turns out that, for each eigenvalue $k_0^2$, there is
is a unique solution with $h$ being regular at $\rho=1$.  
The first few solutions are shown in Table 7.

\begin{table}[htbp]
\centering
\begin{tabular}{l|ccc}
  state & $(-k^2_0)$ & $(- \delta k^2)$ &
$\delta k^2/k^2_0$ \\
 \hline
 $0^{++}$ & 11.59 & 89.75 & 7.74  \\
 $0^{++*}$  & 34.53 & 365.7 & 10.59  \\
 $0^{++**}$  & 68.98 & 809.8 & 11.74 \\
 $0^{++***}$  & 114.9 & 1397 & 12.16 \\
 $0^{++****}$ & 172.3 & 2122 & 12.32 \\
 $0^{++*****}$ & 241.2 & 2991 & 12.40 \\
\end{tabular}
\label{shift}
\parbox{5.5in}{\caption{Leading string correction to the $0^{++}$ glueball
masses in QCD$_3$.
The first column gives the zeroth order supergravity result for the
mass squared, the second column
gives the coefficient of the leading string correction and
the third column gives
their ratio.}}
\end{table}

Recalling that the squared mass of each glueball states is given by
\beq
M^2=- \left( k_0^2 + {1  \over 8} \delta k^2 \zeta(3) \alpha'^3
 + \cdots \right) b^2,
\eeq
we see that the leading stringy corrections to the $0^{++}$
glueball masses are
\begin{eqnarray}
M_{0^{++}}^2 &=& 11.59\times
 (1 +  0.97  \zeta(3) \alpha'^3 + \cdots) b^2 \nonumber \\
M_{0^{++*}}^2 &=& 34.53\times
 (1 +  1.32  \zeta(3) \alpha'^3 + \cdots) b^2\nonumber \\
M_{0^{++**}}^2 &=& 68.98\times
 (1 +  1.47  \zeta(3) \alpha'^3 +\cdots) b^2\nonumber \\
M_{0^{++***}}^2 &=& 114.9\times
 (1 + 1.52  \zeta(3) \alpha'^3 +\cdots) b^2 \nonumber \\
M_{0^{++****}}^2 &=& 172.3\times
 (1 + 1.54  \zeta(3) \alpha'^3 +\cdots) b^2 \nonumber \\
M_{0^{++*****}}^2 &=& 241.2\times
 (1 + 1.55  \zeta(3) \alpha'^3 +\cdots) b^2 ~.
\label{masscorrection}
\end{eqnarray}
It is important to note that the relation between
the radius $R$ of the compactifying circle ${\bf R}^4 \rightarrow
{\bf R}^3 \times {\bf S}^1$ and the location $\rho=b$ of the
horizon also receives an $\alpha'$-correction. Instead of
(\ref{horizonlocation}), we now have \cite{GKT}
\beq
   b = \left(1 - {15 \over 8} \zeta(3) \alpha'^3 + \cdots \right)
{1 \over 2 R}.
\eeq
Therefore, in units of the ultraviolet cutoff $\Lambda_{UV} =
(2R)^{-1}$, the glueball masses are expressed as
\begin{eqnarray}
M_{0^{++}}^2 &=& 11.59\times
 (1 -2.78  \zeta(3) \alpha'^3 + \cdots) \Lambda_{UV}^2 \nonumber \\
M_{0^{++*}}^2 &=& 34.53\times
 (1 -2.43  \zeta(3) \alpha'^3 + \cdots) \Lambda_{UV}^2 \nonumber \\
M_{0^{++**}}^2 &=& 68.98\times
 (1 - 2.28  \zeta(3) \alpha'^3 +\cdots) \Lambda_{UV}^2 \nonumber \\
M_{0^{++***}}^2 &=& 114.9\times
 (1 -2.23  \zeta(3) \alpha'^3 +\cdots) \Lambda_{UV}^2  \nonumber \\
M_{0^{++****}}^2 &=& 172.3\times
 (1 -2.21  \zeta(3) \alpha'^3 +\cdots) \Lambda_{UV}^2  \nonumber \\
M_{0^{++*****}}^2 &=& 241.2\times
 (1 -2.20  \zeta(3) \alpha'^3 +\cdots) \Lambda_{UV}^2 ~.
\label{truecorrection}
\end{eqnarray}

Thus the glueball masses are indeed modified by
the $\alpha' = (4 \pi g_s N)^{-1/2}$ correction.
The corrections are negative for all the 6 levels
we computed and are of the order $\alpha'^3$. Therefore
the glueball masses decrease as we decrease
the `t Hooft coupling $\lambda = g_s N$. As
we discussed in the introduction of this paper, this
is in accordance with our expectation about the continuum
limit of QCD.

At the same time, the $\alpha'$ corrections
to the ratios of the masses appear to be smaller
than the corrections to each glueball mass,
suggesting that the corrections are somewhat universal.
This may indicate that the good agreement between
the supergravity computation and the lattice
gauge theory results is not a coincidence but
is due to small $\lambda^{-1}$ corrections
to the mass ratios.
Obviously, with the given data, we cannot tell
whether the stringy corrections for the mass ratios
remain small in the continuum limit $\lambda \rightarrow 0$. It would
be very interesting to see whether this trend continues
in the subleading corrections in $\alpha'$.

\section{Summary and Discussion}

In this paper, we computed the glueball masses of
large $N$ QCD in three and four dimensions by solving
supergravity wave equations in the AdS black hole geometry.
The supergravity approximation is valid for large
$N$ and large $\lambda= g_{YM}^2 N$ and
therefore the results are to be compared
with a fixed ultraviolet cutoff in the strong ultraviolet coupling
regime.

We computed the ratios of the masses of the excited glueball
states with the mass of the lowest state,
as well as the ratio of masses of two different lowest glueball states.
These ratios are in surprisingly good agreement with
the available lattice data. We also
computed the leading $\lambda^{-3/2}$
corrections to the glueball masses taking into account
stringy corrections to the black hole geometry. We found
that the corrections to the masses are in accordance
with our  expectation about the continuum limit of QCD.
The corrections
to the ratios of the masses appear to be smaller
than the corrections to each glueball mass,
suggesting that the corrections are somewhat universal.

The above computations can be generalized to higher spin glueballs.
As noted previously, the graviton couples to
the energy-momentum tensor and solving its equation of motion
will give the masses of the $2^{++}$ glueball.
In general the higher spin glueballs will correspond
to operators that couple to massive string excitations.
The dimensions of these operators are $\Delta \sim
\lambda^{1/4}$ for large  $\lambda$  \cite{GKP}.
It would be interesting to see how to extrapolate this
result to the continuum $\lambda \rightarrow 0$.

Another interesting issue is the existence of
$SO(6)$ non-singlet states in supergravity. For large
$\lambda$, their masses are of the same order as
the $SO(6)$ singlet states we studied in this paper.
In the continuum limit, $\Lambda_{UV} \rightarrow \infty$
and $\lambda \rightarrow 0$,
those states should decouple.
Presumably $\lambda^{-1}$ corrections make them heavy.

Maldacena's conjecture reduces the problem of solving
large $N$ QCD in three and four dimensions to that of
controlling the $\alpha'$ corrections to the two-dimensional
sigma-model with the Ramond-Ramond background.
In this paper, we have extracted information about
glueballs in strongly coupled QCD using the
$\alpha'$-expansion of the sigma-model. It would certainly
be interesting to understand properties of such a sigma-model better.

\section*{Acknowledgments}

We would like to thank M. Chanowitz, D. Gross, A. Hashimoto,
C. Morningstar,  S. Sharpe and Y. Takei for
discussions and communications. We also thank Harlan Robins and
Jonathan Tannenhauser, who participated in an early stage
of this project, for discussions.
H.O. and Y.O. thank the Institute for Theoretical Physics
at  Santa Barbara for
its hospitality. This work was supported in part by the NSF
grant PHY-95-14797 and in part the DOE grant DE-AC03-76SF00098.
In addition, H.O. and Y.O. are supported by
the  NSF grant PHY-94-07194 through the ITP. C.C. is a research
fellow of the Miller Institute for Basic Research in Science.

\end{document}